\documentclass[a4paper,12pt]{article}

\newcommand{\sect}[1]{\setcounter{equation}{0}\section{#1}}

\newcommand{\bea}{\begin{eqnarray}}
\newcommand{\eea}{\end{eqnarray}}
\newcommand{\be}{\begin{equation}}
\newcommand{\ee}{\end{equation}}
\newcommand{\vs}[1]{\vspace{#1 mm}}

\newcommand{\dsl}{\pa \kern-0.5em /}

\newcommand{\pa}{\partial}

\begin{document}
\topmargin 0pt
\oddsidemargin 0mm

\begin{flushright}
hep-th/0201232\\
SINP-TNP/02-7 
\end{flushright}

\vs{2}
\begin{center}
{\Large \bf  
Correlators Of Open D-string Theory From Supergravity}
\vs{10}

{\large Somdatta Bhattacharya and Shibaji Roy}
\vspace{5mm}

{\em 
 Saha Institute of Nuclear Physics,
 1/AF Bidhannagar, Calcutta-700 064, India\\
E-Mails: som, roy@theory.saha.ernet.in\\}
\end{center}

\vs{5}
\centerline{{\bf{Abstract}}}
\vs{5}
\begin{small}
We consider a minimally coupled massless scalar propagating in the
background of NS5-branes in the presence of a 2-form RR electric 
field. The supergravity solution also called the (NS5,D1) bound state
in the appropriate decoupling limit is the holographic dual of Open
D-string (OD1) theory. Using this information, we compute the two-point
function of the operators in OD1 theory which couples to the massless
string states such as the dilaton. We will indicate how to obtain
the absorption cross-section of a scalar on NS5-branes (i.e. Little
String Theory (LST)) obtained earlier, from our results in the low-energy
limit.
\end{small}

\sect{Introduction}

It is well-known \cite{gmss} from world-volume arguments that in 
six space-time
dimensions there exist a series of new non-gravitational non-local
theories which appear as particular decoupling limits of type IIB
or IIA NS5-branes in the presence of $(p+1)$-form RR electric gauge
fields. As the world-volume theory decouples from the bulk, the 
corresponding\footnote{What we mean by `corresponding' here is that
the $(p+1)$-form gauge field in the supergravity solution becomes the
$(p+1)$-form field-strength on the world-volume by gauge invariance.}
$(p+1)$-form field strength attains a critical value and almost balances
the tension of the D$p$-branes living on the world-volume of NS5-branes.
Thus these theories contain excitations which include fluctuating
light open D$p$-branes and are called OD$p$ theories. The idea for the
existence of such theories came from a series of S- and T-duality
transformations of the underlying string theory starting from Non-Commutative
Yang-Mills (NCYM) theory \cite{sw,hi,mr,aos} in 3+1 dimensions. 
To be precise NCYM appears
as a generalization of the AdS/CFT correspondence \cite{mal,agmoo} on 
the world volume
of D3-branes in the presence of a constant NSNS $B$ field with space-space
components \cite{sw}. By S-duality it has been argued that the strong coupling
limit of NCYM theory does not give rise to a field theory, rather
it gives a non-local theory called a Non-Commutative Open String (NCOS)
theory in 3+1 dimensions \cite{sst,gmms}. By applying T-duality 
on this 3+1 dimensional
NCOS theory one obtains other NCOS theories in various dimensions 
\cite{harone},
where the 2-form field strength corresponding to the NSNS 2-form gauge
field attains a critical value. S-duality on the six-dimensional NCOS
theory then gives rise to six-dimensional OD1 theory, where now the
2-form field strength corresponding to the RR 2-form gauge field takes
the critical value \cite{gmss,hartwo}. By applying T-duality again 
on the OD1-theory
one obtains the series of six-dimensional theories called the OD$p$-theories
\cite{gmss} with coupling constant $G_{o(p)}^{2}$ and length scale 
$\sqrt{\alpha _{\rm eff}^{\prime }}$.
In this paper, we will mainly be interested in OD1 theory, with
coupling $G_{o(1)}^{2}$ and length scale 
$\sqrt{\alpha _{\rm eff}^{\prime }}$.

Holographic dual descriptions of these various world volume theories
are given by the corresponding bound state solutions (having sixteen
super charges) of the supergravity equations of motion with appropriate
decoupling limits. Thus the holographic dual of 3+1 NCYM theory is
the (D1,D3) bound state \cite{bmm,rt,cp} with the NCYM decoupling limit
\cite{sw,hi,mr}. 
Similarly
(F,D$p$) bound states \cite{lurone} with NCOS decoupling limits 
are the holographic
duals of NCOS theories in $(p+1)$ dimensions \cite{harone}. The holographic 
duals
of OD$p$ theories in six dimensions are given by the (NS5,D$p$) bound
state solutions \cite{hartwo,aor,mitro,cgnn} with OD$p$ decoupling limits. 
Since we will be working
with OD1 theory, we will consider only the (NS5,D1) bound state solution
of type IIB string theory.

Since OD$p$ theories are defined on the world-volume of NS5 branes and
it is not known how to quantize 5-branes, very little is known about
the dynamics of such theories. So for example, we only know that OD1
theory has a coupling $G_{o(1)}^{2}$ and length scale 
$\sqrt{\alpha _{\rm eff}^{\prime }}$
and it must reduce to Little String Theory \cite{brs,sei,dvvone,dvvtwo,lms}
at low energies where 
$G_{o(1)}^{2}
 \rightarrow  0$ and $\alpha'_{\rm eff}$ = fixed. So
in the absence of our knowledge about the world volume dynamics, we
will use its holographic description and try to gain some insight
into this theory. The (NS5,D1) bound state solution and the associated
OD1 decoupling limit have been obtained in \cite{gmss,aor,mitro}. We study 
the propagation
of a massless scalar minimally coupled to this background geometry.
We note that the background geometry in the string frame becomes asymptotically
flat, whereas in the near horizon limit it is given by the (NS5,D1)
supergravity metric with the OD1 limit. Unlike in the D3 brane case,
the dilaton is not a constant here and so we cannot trust the (NS5,D1)
supergravity solution all the way through the energy $u  \rightarrow 
0$ in the boundary theory and we have to use the S-dual solution i.e.
the (F,D5) supergravity solution \cite{lurone} in the corresponding 
NCOS limit \cite{harone}.
However, the equation of motion and therefore the solution remains
the same in terms of the dual variables. After finding the unique
solution to the minimally coupled scalar, $\varphi (k,u)$, we
extract the two-point function of the operator ${\cal O}(k)$ 
in the boundary OD1 theory dual to the field $\varphi$ by the
usual procedure \cite{gkp,witt}. To be exact, we identify in the spirit 
of the AdS/CFT
correspondence
\begin{equation}
\left\langle e^{i\int d^{6}k\varphi _{0}(k){\cal O}(k)}\right\rangle =
e^{iS_{\rm sugra}[\varphi (k,u)|_{u=\Lambda }]}
\end{equation}
where on the r.h.s. $S_{\rm sugra}$ represents the on-shell supergravity
action of the scalar $\varphi (k,u)$ in the background geometry
mentioned above. That is on imposing equation of motion, the scalar
action gives
a boundary term which has to be evaluated at the boundary $u = \Lambda$,
where $\varphi (k,u)|_{u=\Lambda }\rightarrow \varphi _{0}(k)$,
with proper renormalization. Since the l.h.s. represents the generating
functional of the operator ${\cal O}(k)$ of the boundary theory, the two point
function can be evaluated from $S_{\rm sugra}$ by taking the functional
derivative with respect to $\varphi _{0}(k)$
twice. We will evaluate the correlation function only in momentum
space as the operators are always well defined in momentum space for
a translationally invariant theory. But as we will see, because of
the momentum dependent renormalization, it is not possible to obtain
the correlator in position space unambiguously \cite{misei,mr} 
which is indicative
of the non-local nature of the boundary theory. We will also point
out how in the low energy limit, our result reproduces the form of
the absorption cross-section of a scalar on NS5-branes \cite{gk,misei}
or Little String
Theory (LST) as expected.

The organization of this paper is as follows. In section 2, we give
the holographic dual description of OD1 theory. The computations of
the boundary action and the two-point function of operators in OD1
theory are given in section 3. Here we also mention how at low energy
we reproduce the absorption cross-section of LST. In section 4 we
present our conclusion.

\sect{Holographic dual of OD1 theory}

As we mentioned in the introduction, the holographic dual description
of OD1 theory is given by the (NS5,D1) supergravity solution in a
particular low energy limit called the OD1 decoupling limit. The string-metric
and the dilaton for the above bound state are given as follows 
\cite{mitro}\footnote{For our purpose here we do not need the other 
fields which can be
found in \cite{mitro}.}, 
\begin{eqnarray}
ds^{2} &=& H^{\frac{1}{2}}H'^{\frac{1}{2}}[H^{-1}(-(dx^{0})^{2}+
(dx^{1})^{2})+H'^{-1}\sum ^{5}_{i=2}(dx^{i})^{2}+dr^{2}+r^{2}
d\Omega ^{2}_{3}]\nonumber\\
e^{\phi } &=& g_{s}H^{\frac{1}{2}}
\end{eqnarray}
 In the above `$r$' is the radial coordinate transverse to the NS5-branes.
The D-strings lie along the $x^{1}$ direction. Also, $d\Omega ^{2}_{3}
= d\theta ^{2}+\sin^{2}\theta d\phi ^{2}_{1}+\sin^{2}\theta \sin^{2}\phi _{1}
d\phi ^{2}_{2}$
is the line element of the unit sphere transverse to NS5-branes. $g_{s} = 
e^{\phi _{0}}$
is the string coupling constant. The harmonic functions have the forms
\begin{eqnarray}
H &=& 1+\frac{Q_{5}}{r^{2}}\nonumber\\
H^{\prime } &=& 1+\frac{\cos^{2}\psi Q_{5}}{r^{2}}
\end{eqnarray}
where the charge $Q_{5}$ and the angle $\cos\psi$ are given
as follows,
\begin{eqnarray}
Q_{5} &=& \frac{N\alpha ^{\prime }}{\cos\psi }\nonumber\\
\cos\psi &=& \frac{N}{\sqrt{M^{2}g_{s}^{2}+N^{2}}}
\end{eqnarray}
with $N$, the number of NS5-branes and $M$, the number of D-strings per
$(2\pi )^{4}\alpha ^{\prime 2}$ of four co-dimensional area of
NS5-branes. The OD1 decoupling limit \cite{gmss,aor,mitro} 
is obtained by defining a dimensionless
scaling parameter $\epsilon$ and taking $\cos\psi =\epsilon \rightarrow 0$,
keeping the following quantities fixed,
\begin{equation}
\alpha ^{\prime }_{\rm eff} = \frac{\alpha ^{\prime }}{\epsilon },\qquad 
u = \frac{r}{\epsilon \alpha ^{\prime }_{\rm eff}},\qquad G_{o(1)}^{2}
= \frac{g_{s}}{\epsilon },\qquad Q_{5}=\alpha ^{\prime }_{\rm eff}N
\end{equation}
With these the harmonic functions in (2.2) take the forms,
\begin{eqnarray}
H &=& \frac{1}{\epsilon ^{2}a^{2}u^{2}}\nonumber\\
H^{\prime } &=& \frac{h}{a^{2}u^{2}}
\end{eqnarray}
where $h = 1 + a^2 u^2$, with $a^2 = \alpha'_{\rm eff}/N$ 
and the metric and the dilaton reduce to
\begin{eqnarray}
ds^{2} &=& \alpha ^{\prime }h^{\frac{1}{2}}[-(d\tilde{x}^{0})^{2}+
(d\tilde{x}^{1})^{2}+h^{-1}\sum ^{5}_{i=2}(d\tilde{x}^{i})^{2}+
\frac{N}{u^{2}}(du^{2}+u^{2}d\Omega _{3}^{2})]\nonumber\\
e^{\phi } &=& \frac{G_{o(1)}^{2}}{au}
\end{eqnarray}
In the above \( \sqrt{\alpha ^{\prime }_{\rm eff}} \) is the length scale,
$u$ is the energy parameter and \( G_{o(1)}^{2} \) is the coupling
constant of OD1 theory. Note that even though \( u \) is fixed, it
can be set to an arbitrarily small or large values corresponding to
the IR or UV limit of OD1 theory. The fixed coordinates are defined
as

\begin{equation}
\tilde{x}^{0,1}=\frac{1}{\sqrt{\alpha ^{\prime }_{\rm eff}}}x^{0,1},\qquad 
\tilde{x}^{2,\ldots,5}=\frac{1}{\epsilon \sqrt{\alpha ^{\prime }_{\rm eff}}}
x^{2,\ldots,5}
\end{equation}
Eq.(2.6) represents the supergravity dual of OD1-theory.

The supergravity
description remains valid as long as the curvature measured in units
of \( \alpha ^{\prime } \) i.e. \( \alpha ^{\prime } {\cal R} =
1/(1+a^{2}u^{2})^{\frac{1}{2}}N \ll 1 \)
and the dilaton \( e^{\phi }= G_{o(1)}^{2}/(au) \ll 1. \) This
can be satisfied if $N$ is large and \( au \gg G^{2}_{o(1)}. \) So in
the extreme IR region \( (u\rightarrow 0), \) we cannot trust the
supergravity solution (2.6) and we have to use its S-dual version i.e.
the (F,D5) bound state with the corresponding NCOS decoupling limit 
\cite{cmor}.
But we will point out later that this will not change the results.

We would like to mention that in the low-energy limit  
we would expect that the
OD1 theory would reduce to LST \cite{gmss}. However, this is not quite
transparent from the OD1 limit in (2.4) and the supergravity configuration
(2.6). In fact, the OD1-limit looks quite different from the LST limit
\( g_{s}\rightarrow 0 \) and \( \alpha ^{\prime } \) = fixed on
NS5-branes as has been also pointed out in refs.\cite{aor,mitro}. The reason
can be understood from the angle \( \cos\psi  \) given in (2.3). Notice
here that since \( \cos\psi =\epsilon \rightarrow 0, \) the number
of D-strings is very large compared to the number of NS5-branes. Since
D-strings dominate over NS5-branes, it is not clear how to obtain
the NS5-brane world-volume theory only, in the OD1 limit. Even from
the metric in (2.6), although in the low-energy limit \( h\approx 1, \)
and we recover six dimensional Poincare invariant theory, it does
not look like the holographic dual of LST, because the metric is multiplied
with \( \alpha ^{\prime }\rightarrow 0 \) (instead of some fixed
\( \alpha ^{\prime }). \) A decoupling limit different from (2.4) which
closely resembles the LST limit has been discussed in \cite{hartwo}. This
limit has been referred to as the OBLST limit and from here it is
easy to recover the holographic dual of LST in the low-energy limit.
But this type of limit holds only for (NS5,D1) and (NS5,D2) and cannot
be generalised for other (NS5,D$p$) cases. This is the reason we work
with the OD1-limit (2.4).

The clue to recover LST from (2.6) is the fact that we have to take
\( G_{o(1)}^{2}\rightarrow 0 \) with \( \alpha ^{\prime }_{\rm eff} \)
fixed. To see precisely how this can be achieved one should look closely
at the angle given in (2.3). Notice that since \( \cos\psi =\epsilon 
\rightarrow 0, \)
it implies that \( N/Mg_{s} = N/(MG^{2}_{o(1)}\epsilon ) \sim \epsilon . \)
So if \( G^{2}_{o(1)} \) = fixed, then \( N/M \sim \epsilon ^{2} \)
; on the other hand, if \( G_{o(1)}^{2}\sim \epsilon \rightarrow 0, \)
such that \( \cos\psi \approx 1 \), then both \( H \) and \( H^{\prime } \)
in (2.2) are the same and have the values \( H = H^{\prime }=1/(a^{2}
\epsilon ^{2}u^{2}). \)
Using this and \( u= r/(\epsilon \alpha ^{\prime }_{\rm eff}), \)
it can be easily checked that (2.1) indeed reduces to the holographic
dual of LST \cite{imsy}. Note that in this case 
\( \alpha '=\alpha '_{\rm eff} \)
= fixed. The parameter \( \epsilon  \) is then defined by the relation
\( u = r/(\epsilon \alpha '_{\rm eff}). \) Thus we conclude that
LST is the low-energy limit of OD1 theory. Since (5+1) dimensional
NCOS theory is S-dual to OD1 theory, the coupling constant and the
length scale \( G_{o}^{2} \) and \( \sqrt{\tilde{\alpha }^{\prime }_{\rm eff}}
 \)
of NCOS theory are related to those of OD1 theory by \( G^{2}_{o} = 
1/G_{o(1)}^{2} \)
and \( \sqrt{\tilde{\alpha }^{\prime }_{\rm eff}} \) =
\( G_{o(1)}\sqrt{\alpha ^{\prime }_{\rm eff}}. \)
So in order to recover LST from NCOS theory we should take 
\( G_{o}^{2}\rightarrow \infty  \)
and \( \tilde{\alpha }^{\prime }_{\rm eff}\rightarrow 0, \) such that
\( G_{o}^{2}\tilde{\alpha }^{\prime }_{\rm eff} \) =
\( \alpha ^{\prime }_{\rm eff} \)
= fixed. One can check by a similar argument that LST can indeed be
obtained in this limit.

In the next section when we compute the correlator of OD1 theory we
will point out how to obtain the absorption cross-section of a minimally
coupled massless scalar on NS5-branes or LST obtained before \cite{misei,gk},
 from
our result in the low energy limit.

\sect{Boundary action and the two point function of OD1 theory}

Here we consider the propagation of a minimally coupled massless scalar
\( \varphi  \) in the background geometry of the (NS5,D1) bound state
solution given in eq.(2.1) in the corresponding decoupling limit. The
scalar \( \varphi  \) corresponds to the fluctuation of the dilaton
(or it could be certain longitudinal components of the graviton).
Setting all other fluctuations to zero, the action for the scalar
can be written as,
\begin{equation}
S_{\rm sugra}=\frac{1}{(2\pi )^{7}\alpha ^{\prime 4}}
\int d^{10}x\sqrt{-g}e^{-2\phi }[\frac{1}{2}g^{\mu \nu }\partial _{\mu }
\varphi \partial _{\nu }\varphi ]
\end{equation}
We have written the action in the string frame. Using the background
geometry given in (2.1) and assuming that \( \varphi  \) is in the
$s$-wave, the above action can be written as follows,
\begin{equation}
S_{\rm sugra}=\frac{1}{4(2\pi )^{5}g_{s}^{2}\alpha '^{4}}
\int d^{6}x\int drr^{3}[(\partial _{r}\varphi )^{2}+H\eta ^{\mu \nu }
\partial _{\mu }\varphi \partial _{\nu }\varphi +
H'\delta ^{ij}\partial _{i}\varphi \partial _{j}\varphi ]
\end{equation}
where in (3.2) we have performed the integration over the angular part
yielding a factor \( \Omega _{3}=2\pi ^{2}. \) The indices \( \mu \,, \nu =
0,1 \)
and \( i,j=2,\ldots,5. \) \( H \) and \( H' \) are harmonic functions
given in (2.2). The equation of motion following from (3.2) is
\begin{equation}
[\frac{1}{r^{3}}\partial _{r}(r^{3}\partial _{r})+H\eta ^{\mu \nu }
\partial _{\mu }\partial _{\nu }+
H'\delta ^{ij}\partial _{i}\partial _{j}]\varphi =0
\end{equation}
In the OD1-decoupling limit (2.4), the above equation reduces to 
\begin{equation}
[\frac{1}{u^{3}}\partial _{u}(u^{3}\partial _{u})-
\frac{Nk^{2}}{u^{2}}-k'^{2}]\tilde{\varphi }(u)=0
\end{equation}
where,
\begin{equation}
\varphi (x^{\mu },r)=e^{ik\cdot x}\tilde{\varphi }(u)\varphi _{0}(k)
\end{equation}
with \( \mu =0,1,\ldots,5 \) and \( k\cdot x=\sum _{\mu =0}^{5}
k_{\mu }x^{\mu } \).
Also in (3.4) we have defined \( k^{2}=\sum _{i=1}^{5}
\tilde{k}^{2}_{i}-\tilde{\omega }^{2} \)
and \( k'^{2}=\alpha '_{\rm eff}\sum _{j=2}^{5}\tilde{k}_{j}^{2}. \) 

The `tilde' variables are the scaled variables and are related to
the original variables as \( \tilde{\omega }=\sqrt{\alpha '_{\rm eff}}\omega
 \),
\( \tilde{k}_{1}=\sqrt{\alpha '_{\rm eff}}k_{1} \) and \( \tilde{k}_{j}=
\epsilon \sqrt{\alpha '_{\rm eff}}k_{j} \),
for \( j=2,\ldots,5. \) After defining the scaled variables this way,
we will drop the `tildes' for notational simplicity from now on. We
recognize eq.(3.4) to be the equation for the modified Bessel function
with two independent solutions given by 
\begin{equation}
\tilde{\varphi }(u)=u^{-1}I_{\nu }(k'u)\qquad {\rm or}\qquad 
u^{-1}K_{\nu }(k'u)
\end{equation}
where \( k'=\sqrt{k'^{2}} \) and \( \nu ^{2}=1+Nk^{2}, \) with \( I_{\nu } \)
and \( K_{\nu } \) being the modified Bessel functions. From these
the solution which will be chosen must be well behaved as \( u\rightarrow 0. \)
So, the unique solution that is picked would be,
\begin{equation}
\tilde{\varphi }(u)=u^{-1}I_{\nu }(k'u)
\end{equation}

A word of caution is needed at this point. We have mentioned earlier
that the (NS5,D1) supergravity solution may not be trustworthy at
arbitrarily small values of the energy parameter, since it may violate
the condition \( e^{\phi }=G_{o(1)}^{2}/(au)\ll 1. \) In that case one
has to use the S-dual configuration i.e. (F,D5) solution with the
corresponding NCOS limit. But one can check that this will lead to
the same equation of motion giving the same solution as in the (NS5,D1)
case. Now, using the following identity,
\begin{equation}
I_{\nu }(k'u)=e^{-i\nu \pi /2}J_{\nu }(ik'u)
\end{equation}
where \( J_{\nu } \) is the ordinary Bessel function, we obtain the
following asymptotic behaviour of \( \tilde{\varphi }(u) \),
\begin{equation}
\tilde{\varphi }(u)\rightarrow \frac{1}{\sqrt{2\pi k'}u^{\frac{3}{2}}}
[e^{-k'u}e^{-i(\nu +\frac{1}{2})\pi }+e^{k'u}]
\end{equation}
for \( u\rightarrow \infty . \) Note here that the coefficient of
\( e^{-k'u} \) in (3.9) is like the reflection coefficient for the
scalar. However, we would like to point out that since \( k' \) does
not include the time-like momenta, so this is not really the reflection
coefficient. If on the other hand, we had set all the space-like momenta
to zero, then the equation of motion (3.4) would reduce to 
\begin{equation}
[\frac{1}{u^{3}}\partial _{u}(u^{3}\partial _{u})+
\frac{N\tilde{\omega }^{2}}{u^{2}}]\tilde{\varphi }(u)=0
\end{equation}

The solution for this equation has the form
\begin{equation}
\tilde{\varphi }(u)=u^{-1\pm \sqrt{1-N\tilde{\omega }^{2}}}
\end{equation}
So in this case there will be non-trivial absorption if 
\begin{equation}
\tilde{\omega }>\frac{1}{\sqrt{N}}\qquad {\rm or} \qquad 
\omega >\frac{1}{\sqrt{\alpha '_{\rm eff}N}}
\end{equation}
This is similar to the mass gap found for LST \cite{misei,ali}.

Now in order to evaluate the two-point function, it has to be properly
renormalised. We here adopt the technique used in \cite{misei,mr} for this
purpose. Defining \( u=\Lambda  \) (where \( \Lambda  \) is the
UV cutoff of the OD1 theory) as the boundary, we demand the following
boundary condition for our solution \( \tilde{\varphi }(u) \),
\begin{equation}
\tilde{\varphi }(\Lambda )=\frac{1}{\sqrt{2\pi k'}}\frac{1}
{\Lambda ^{\frac{3}{2}}}e^{k'\Lambda }\qquad {\rm as}\quad 
\Lambda \rightarrow \infty 
\end{equation}
This condition ensures that the solution remains finite in the interior
as \( \Lambda \rightarrow \infty  \). We are now in a position to
evaluate the boundary action. Integrating by parts and using (3.3),
we obtain from (3.2)
\begin{eqnarray}
S_{\rm sugra} &=& \frac{1}{2(2\pi )^{5}g_{s}^{2}\alpha '^{4}}
\int d^{6}x\int dr\partial _{r}(r^{3}\varphi \partial _{r}\varphi )\nonumber\\
&=& \frac{\alpha '_{\rm eff}}{2(2\pi )^{5}G^{4}_{o(1)}}
\int d^{6}x\int du\partial _{u}(u^{3}\varphi \partial _{u}\varphi )
\end{eqnarray}
Expanding \( \varphi (x^{\mu }) \) as 
\begin{equation}
\varphi (x^{\mu })=\int d^{6}ke^{ik\cdot x}\varphi _{0}(k)
\end{equation}
eq.(3.14) can be simplified to give 
\begin{equation}
S_{\rm sugra}=\frac{\pi \alpha '_{\rm eff}}{G_{o(1)}^{4}}
\int d^{6}kd^{6}q\delta ^{6}(k+q)\varphi _{0}(k)\varphi _{0}(q){\cal F}
\end{equation}
The flux factor ${\cal F}$ is defined as 
\begin{equation}
{\cal F}=\tilde{\varphi }(u)u^{3}\partial _{u}
\tilde{\varphi }(u)]_{u=0}^{u=\Lambda }
\end{equation}
Since \( \varphi  \) falls off exponentially as \( u\rightarrow 0, \)
we need to evaluate (3.17) for \( u=\Lambda  \) (cut-off surface).
Using (3.9) and (3.13) in (3.16), we obtain the supergravity action on
the boundary as,
\begin{equation}
S_{\rm sugra}=\frac{\pi \alpha '_{\rm eff}}{G_{o(1)}^{4}}
\int d^{6}k[div(\Lambda ,k)-\frac{1}{2\pi }e^{-i(\nu +\frac{1}{2})\pi }+....]
\varphi _{0}(k)\varphi _{0}(-k)
\end{equation}
The first term is divergent as \( \Lambda \rightarrow \infty  \)
and should be subtracted away. By the `dots' we indicate terms that
vanish as \( \Lambda \rightarrow \infty . \) So the two point function
has the form\begin{equation}
\left\langle {\cal O}(k){\cal O}(-p)\right\rangle \sim 
\frac{\alpha '_{\rm eff}}{2G_{o(1)}^{4}} e^{-i\nu \pi}\delta^{6}(k-p)
\end{equation}
It is clear that if $\nu$ is real i.e. if $1+N(\sum_{i=1}^5 k_i^2 - \omega^2)
> 0$, then the correlation function is trivial in the sense that it becomes
an analytic function of the momentum. On the other hand if we define
$[1+N(\sum_{i=1}^5 k_i^2 - \omega^2)]^{1/2} = - i\mu$, for $\mu$ = real, then
the correlation function is non-trivial and has the form,
\begin{equation}
\langle {\cal O}(k) {\cal O}(-p)\rangle \sim \frac{\alpha'_{\rm eff}}
{2 G_{o(1)}^4} e^{-\mu\pi}\delta^6(k-p)
\end{equation}
This is our
result for the two-point function of an operator in the OD1 theory
which couples to the massless scalar in supergravity. For small values
of \( \mu , \) eq.(3.20) reduces to \begin{equation}
\left\langle {\cal O}(k){\cal O}(-p)\right\rangle \sim \frac{\alpha '_{\rm eff}
\pi }{2G_{o(1)}^{4}}[N(\omega^2 - \sum_{i=1}^5 k_i^2) - 1]^{1/2}
\delta ^{6}(k-p)
\end{equation}
Note that even if there is no real absorption here (see eq.(3.9) and the
remarks thereafter), we still get a non-trivial correlation function.

Now we will show how the absorption cross-section of a scalar
on the NS5-brane theory obtained before \cite{gk,misei} 
can be deduced in the low-momentum
limit. Note that for low momentum we can rewrite the equation of motion
(3.4) as,
\begin{equation}
[\frac{1}{u^{3}}\partial _{u}(u^{3}\partial _{u})-
\frac{Nk^{2}}{u^{2}}-k^{2}]\tilde{\varphi }(u)=0
\end{equation}
The solution for this equation would be given as \begin{equation}
\tilde{\varphi }(u)=u^{-1}I_{\nu }(ku)
\end{equation}
The asymptotic behaviour of this function is\footnote{For large $u$, we
should not neglect 1 in both the harmonic functions $H$ and $H'$ in
(2.2). Then it can be checked from (3.3) that the equation of motion
has the form (3.22) even for large $u$.} 
\begin{equation}
\tilde{\varphi }(u)\rightarrow \frac{1}{\sqrt{2\pi k}u^{\frac{3}{2}}}
[e^{-ku}e^{-i(\nu +\frac{1}{2})\pi }+e^{ku}]
\end{equation}
Unlike in the case (3.9), $k$ now includes the time-like momentum and
the coefficient of \( e^{-ku} \) is truly the reflection coefficient
of a scalar. So if \( \nu  \) is real then \( |R|^{2}=1 \) and there
is no absorption. In the absence of longitudinal momenta, the reflection
coefficient would be given as\begin{equation}
|R|^{2}=e^{-2\pi \mu }
\end{equation}
where \( \mu =\sqrt{N\alpha '\omega ^{2}-1}. \) 

Therefore, the absorption cross-section would be proportional to 
\begin{equation}
\sigma \sim (1-e^{-2\pi \mu })
\end{equation}
This result was obtained in refs.\cite{gk,misei}. We have thus recovered the
absorption cross-section of a scalar on the NS5-brane or LST in the
low-energy limit. Note that since it is the coefficient of $e^{-ku}$ which
determines the form of the correlation function and they have the same
exponential form in both OD1 theory and NS5 brane theory, the correlation
function will have the same structure in this case as in (3.20). But of
course $\mu$ is different in this case and is given by $\mu = [N\alpha'
(\omega^2 - \sum_{i=1}^5 k_i^2) - 1]^{1/2}$ with $\alpha'$ fixed, whereas
for OD1 theory $\mu = [N(\tilde{\omega}^2 - \sum_{i=1}^5 
\tilde{k}_i^2) - 1]^{1/2}$, where the scaled variables were defined before.

It should be emphasized as noted in ref.\cite{misei,mr} that in order to get
the cut-off independent correlation function we had to perform a momentum
dependent renormalization. Because of this, it is not possible to
Fourier transform the momentum-space correlation function to position
space in an unambiguous manner. This is related to the non-local nature
of the boundary theory in the UV.

It is not difficult to generalize our result for the higher partial
waves. The $s$-wave equation of motion (3.3), should be modified for
the $l$-th partial waves as,\begin{equation}
[\frac{1}{r^{3}}\partial _{r}(r^{3}\partial _{r})+
H\eta ^{\mu \nu }\partial _{\mu }\partial _{\nu }+H'\delta ^{ij}
\partial _{i}\partial _{j}-\frac{l(l+2)}{r^{2}}]\varphi =0
\end{equation}
In the OD1-decoupling limit (2.4), it reduces to \[
[\frac{1}{u^{3}}\partial _{u}(u^{3}\partial _{u})-
\frac{Nk^{2}}{u^{2}}-k'^{2}-\frac{l(l+2)}{u^{2}}]\tilde{\varphi }(u)=0\]
Following the same technique as before we obtain the form of the two-point
function as
\begin{equation}
\left\langle {\cal O}(k){\cal O}(-p)\right\rangle \sim 
\frac{\alpha '_{\rm eff}}{2G_{o(1)}^{4}}e^{-\mu \pi} \delta ^{6}(k-p)
\end{equation}
where now,
\begin{equation}
\mu =[N(\omega^2 - \sum_{i=1}^5 k_i^2) - (l+1)^2]^{\frac{1}{2}}
\end{equation}
Note here that $\omega$ and $k_i$'s are scaled variables as in eq.(3.21).
This concludes our calculation of two point function for the operators in OD1
theory which couple to the massless string states.

\sect{Conclusion}

With very little knowledge at hand about the world-volume theory of
NS5-brane in the presence of a near critical RR 2-form electric field,
or OD1 theory, we have taken recourse to its holographic description
to gain some insight into this theory. The (NS5,D1) bound state solution
of type IIB supergravity in a particular low energy limit called the OD1
decoupling limit is the holographic description of OD1 theory. We have
studied the propagation of a massless scalar minimally coupled to this
background. We computed the boundary action and in the spirit of AdS/CFT
correspondence we obtained the form of two point function of the operators
in OD1 theory which couples to massless string states such as the dilaton.
In order to get a finite correlation function, we had to perform a momentum
dependent renormalization for the scalar $\varphi$ and this is the reason we
can not Fourier transform the momentum space correlation function into
position space in an unambiguous way. This is a reflection of the fact
that the boundary theory is non-local. We have also pointed out how to 
obtain the absorption cross-section of a scalar on NS5-branes from our 
results in the low energy limit. Finally, we have extended our results for the
higher partial waves.

In the case of coincident D3-branes, coupling this background to massive
string states one can compute the correlation function of the operators
in the Yang-Mills theory which couple to massive string modes. The position
space correlation function then determines the anomalous dimensions of
the operators in Yang-Mills theory \cite{gkp,witt}. For the case of (NS5,D1)
background also, it is possible to couple it to massive string states.
However, we have not been able to solve the corresponding equation of motion.
In any case, since the boundary OD1 theory is non-local, it will not be
possible to obtain the position space correlation function in an unambiguous
way and hence the determination of the dimension of such an operator would
also be ambiguous.

We have not studied in this paper the structure of the correlation function
for other OD$p$ theories. For odd $p$, the corresponding holographic
description would be given by the type IIB supergravity background in the
appropriate decoupling limits. So, the correlation function in this case
will have similar structure as found in this paper. However, for $p$ = even,
the holographic description would be given by type IIA supergravity background
and for $u \rightarrow 0$, the appropriate description would be given by
uplifting the solution to M-theory. It would be interesting to understand
the structure of correlation function in this case which in the low energy
limit should reproduce the form of the correlation function of LST obtained
in ref.\cite{misei}. It would also be interesting to understand the results
of this paper from the world volume point of view along the line 
\cite{abks,egkrs}.

\end{document}